\documentclass[a4paper,twocolumn]{article}
\usepackage{graphicx}
\begin{document}
\title{Rubidium Spectroscopy at 778-780 nm with a Distributed Feedback Laser Diode}
\author{Sebastian Kraft(1),  Anselm Deninger(2), Christian
Tr\"uck(1),\\
J\'ozsef Fort\'agh(1), Frank Lison(2) and Claus Zimmermann(1)\\[1ex]
(1)Physikalisches Institut, Universit\"at T\"ubingen,\\
Auf der Morgenstelle 14, 72076 T\"ubingen, Germany \\
(2)TOPTICA Photonics AG, Fraunhoferstra{\ss}e 14,\\ 82152 Martinsried,
Germany }
\date{}
\maketitle
We have performed high resolution spectroscopy of rubidium with a
single mode continuous wave distributed feedback (DFB) laser diode.
The saturation spectrum of the D$_2$-line of $^{85}\mathrm{Rb}$ and
$^{87}\mathrm{Rb}$ was recorded with a resolution close to the
natural line width. The emission frequency was actively stabilized
to Doppler-free transitions with a relative accuracy of better than
7 parts in $10^9$ using commercially available servo devices only.
An output power of 80 mW was sufficient to allow for two-photon
spectroscopy of the 5S-5D-transition of $^{87}\mathrm{Rb}$. Further,
we report on the spectral properties of the DFB diode, its tuning
range and its frequency modulation properties. The line width of the
diode laser, determined with high resolution Doppler free two photon
spectroscopy, was 4 MHz without applying any active stabilization
techniques. For time scales below 5 $\mu$s the line width drops
below 2 MHz.\\[2ex]
\hrule \vspace*{5ex}

The commercial introduction of single mode quantum well laser diodes
about 15 years ago has revolutionized high resolution spectroscopy
of atoms and molecules. While the spectral properties and the
tunability of the pure diode chip are not sufficient for many
applications, a reliable and useful device can be constructed by
stabilizing the diode with optical feedback from an external grating
\cite{wieman1991,macadam1992,ricci1995}. Today, so-called external
cavity diode lasers (ECDLs) are commercially available and widely
used in numerous experiments in the fields of quantum optics,
ultra-cold atomic and molecular quantum gases, high resolution
spectroscopy, and metrology. Although highly superior to
conventional laser systems in many aspects, diode lasers still
require considerable expertise for alignment, frequency control and
operation such that the optical apparatus is still a critical part
in a typical experiment.  In particular, for mode-hop free tuning
over a large frequency range, the Bragg angle of the grating and the
resonator length need to be adjusted simultaneously to keep a
longitudinal mode resonant within the laser cavity. Although
mode-hop free tuning ranges in the order of 100 GHz have been
demonstrated with commercially available ECDLs, the precise
mechanical realization of the pivot axis remains a challenge in
terms of engineering, adjustment and stability.

A semiconductor laser that offers the tunability of an
external-cavity system without its mechanical complexity is a
Distributed Feed-Back (DFB) diode \cite{tohmori1993}. In a DFB
diode, a Bragg grating is integrated into the active section of the
semiconductor. Wavelength tuning is realized by altering the
refractive index of the semiconductor, which can be achieved by
either changing the temperature or the operating current. This type
of laser has long been used for telecommunication and can be
regarded as the optical analogue of a voltage controlled oscillator
in radio frequency technology: the emitted wavelength depends only
on well controlled parameters in a robust and reproducible way.
Whilst DFB diodes at 1550 nm proved suitable for spectroscopy of
hydrocarbons \cite{labachelerie1994}, spectroscopic measurements
with alkali atoms have hitherto required second harmonic generation
\cite{poulin1994}. Recently, DFB laser chips have become available
at various wavelengths near the D$_1$- and D$_2$-lines of alkali
atoms. In this article we present a detailed test of a DFB laser for
spectroscopy of the rubidium D$_2$-line and the two-photon 5S-5D
transition. It turns out that with low noise driving electronics its
performance equals that of grating stabilized diode lasers in almost
all aspects. By reducing the complexity of optical experiments, this
diode laser opens new possibilities for upcoming projects that
require automated operation, e.g. optical clocks in satellites or
air-borne experiments in atmospheric physics.

For the measurements presented here, a DL 100 laser system and
control electronics from TOPTICA Photonics were used. Experiments
were carried out with two different DFB diodes with emission ranges
of 779.1-781.0 nm and 777.8 nm-780.3 nm, respectively. The diodes
were mounted into a \lq ColdPack\rq\ housing \cite{bickleder}, a
TO-3 style package with integrated thermistor and four
thermoelectric coolers. The ColdPack permitted rapid and flexible
control of the diode temperature within a range of -10$\dots$+40
$^\circ$C and sweep times of up to 3.5 K/s. A \lq Bias-Tee\rq\ was
included in the laser head to allow for radio frequency modulation
of the laser current and the generation of sidebands within the
laser spectrum. For frequency stabilization, a lock-in regulator was
employed. The laser beam was collimated by an aspheric lens (NA =
0.55). The laser beam profile was shaped from elliptical to circular
by an anamorphic prism pair that compressed the long axis by a
factor of three. A 60 dB optical isolator prevented optical feedback
into the laser diode.

The coarse spectral properties of the DFB laser were characterized
with a commercial grating spectrometer (AQ-6315A Optical Spectrum
Analyzer, Ando Electric) and a high precision wavelength meter
(HighFinesse-Angstrom WS/Ultimate, absolute accuracy 30 MHz).

\begin{figure}
\includegraphics[width=\columnwidth]{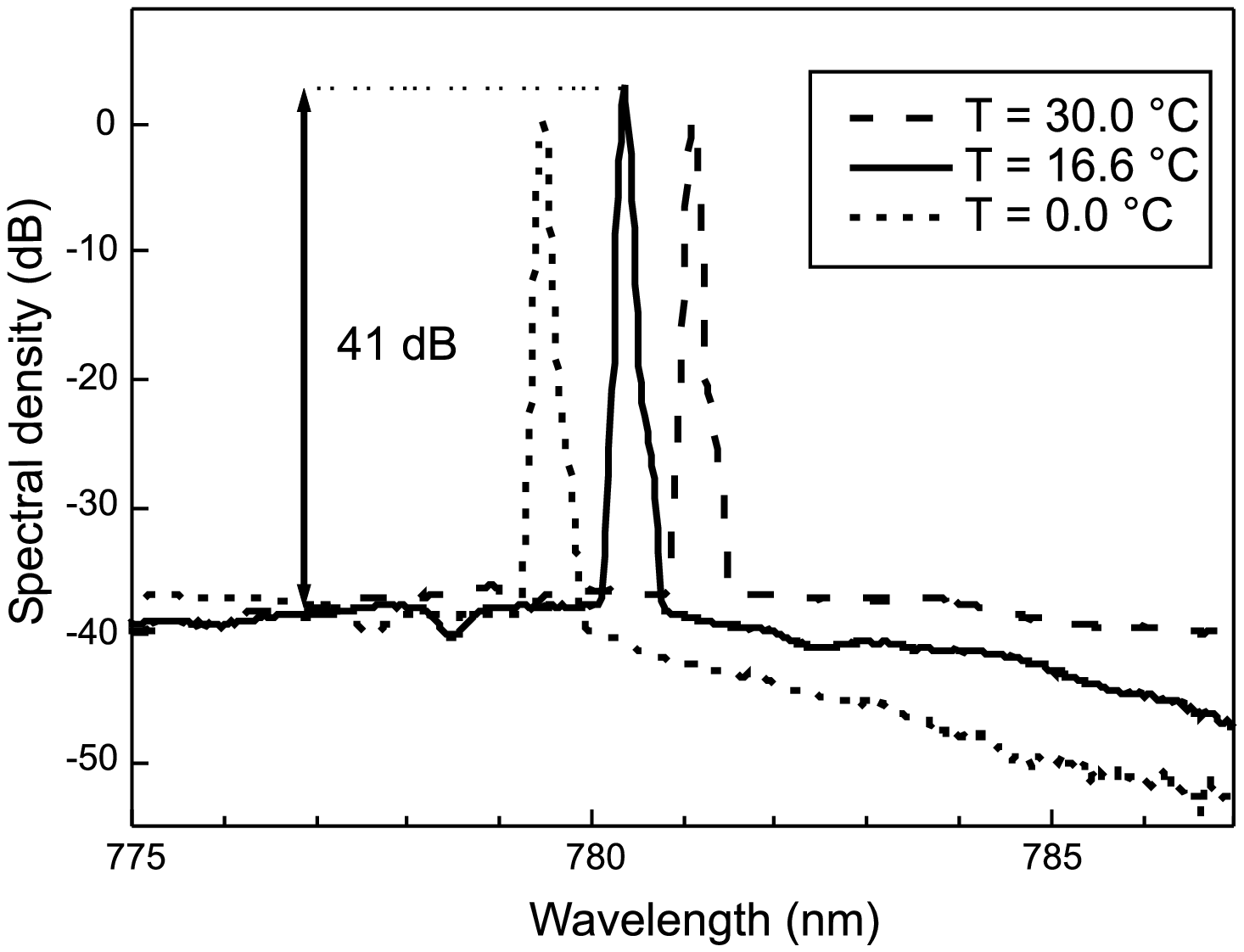}
\caption{Emission spectra of the DFB laser at different diode
temperatures of  0.0 $^\circ $C, 16.6 $^\circ $C and 30 $^\circ $C
(from left to right). The resolution of the grating spectrometer was
0.05 nm.} \label{spectrum}
\end{figure}

Fig. \ref{spectrum} shows the spectrum of the DFB diode at
different
temperatures as measured with the grating spectrometer at 50 pm
resolution. It reveals a background of amplified spontaneous
emission (ASE) which is suppressed by $37\dots41$ dB relative to
the coherent part of the diode radiation. At low temperatures, a
slight increase in ASE is observed on the short-wavelength side of
the spectrum. This results from the difference in the temperature
induced frequency shift of the internal grating, which determines
the lasing wavelength, and of the gain profile of the
semiconductor which is responsible for the ASE. More specifically,
the resonance wavelength of the grating changes by approximately
0.05 nm/K (see below), while the gain spectrum varies at a rate of
$\sim 0.2$ nm/K. Thus, the resonant wavelength shifts to the side
of the gain spectrum, which eventually limits the accessible
wavelength range of the DFB diode.

\begin{figure}
\includegraphics[width=\columnwidth]{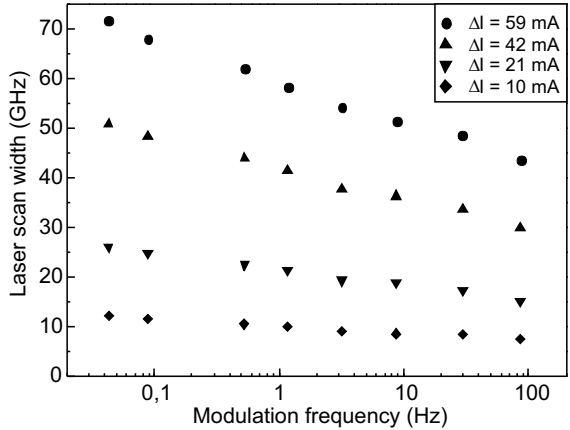}
\caption{Tuning range of the DFB diode with current modulation. The
graph shows the measured scan width as a function of the tuning
speed (modulation frequency in Hz). The symbols represent current
modulation amplitudes of $\Delta$I = 10 mA, $\Delta$I = 21 mA,
$\Delta$I = 42 mA, and $\Delta$I = 59 mA, respectively. }
\label{current-freq}
\end{figure}

The tuning range of the laser was determined with the wavelength
meter. The output frequency can either be changed by varying the
operating current of the diode or the temperature of the chip
housing. The modulation of the current changes both the carrier
density and the internal temperature of the semiconductor chip. The
thermal effect is comparably slow and becomes less relevant with
increasing tuning rates. This is shown in Fig. \ref{current-freq},
which plots the scan width (i.e. the shift of
the emission frequency) as a function of tuning speed. The different
symbols refer to modulation amplitudes of 59 mA, 42 mA, 21 mA and 10
mA, respectively. All measurements indicate an approximately
logarithmic reduction of the scan width with increasing modulation
frequency.  More specifically, the tuning rate [GHz/mA] amounts to
$\approx$ 1.2 GHz/mA at slow modulation frequencies ($f = 42$ mHz)
and decreases to $\approx$ 0.7  GHz/mA at faster modulation
frequencies ($f = 86$ Hz). Moreover, the tuning rate is independent
of the modulation amplitude, i.e. at any fixed modulation frequency,
the ratio of the scan widths in Fig. \ref{current-freq} reproduces
precisely the ratio of the different modulation amplitudes employed.
Generally, the comparatively large tuning rates call for a low-noise
current driver for high resolution applications with laser line
widths in the 1 MHz range.

\begin{figure}
 \includegraphics[width=\columnwidth]{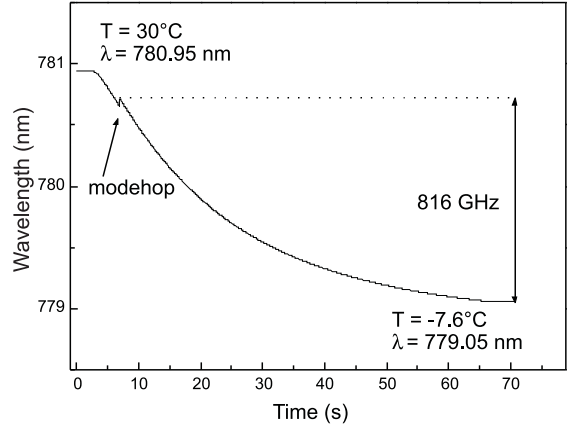}
\caption{Thermal tuning characteristics of the DFB diode. Shown is
the change of the emission wavelength while the diode was cooled
from 30 $^\circ $C to -7.6 $^\circ $C. The mean thermal tuning
rate is $\Delta \nu $ / $\Delta $T = - 25 GHz / K.}
\label{temp}
\end{figure}

The thermal tuning characteristics are illustrated in Fig.
\ref{temp}. The diode temperature was
decreased from 30 $^\circ $C to -7.6 $^\circ $C while the emission
wavelength was monitored with the wavelength meter. The wavelength
varied between 780.95 nm and 779.05 nm, with one mode-hop by 0.06
nm (30 GHz) occurring at 780.7 nm. Single-mode emission was
maintained over a wavelength range of more than 1.6 nm (814 GHz).
The data yield an average thermal tuning rate of $\Delta \nu$ /
$\Delta$T = - 25 GHz / K (+ 0.05 nm / K).

\begin{figure}
 \includegraphics[width=\columnwidth]{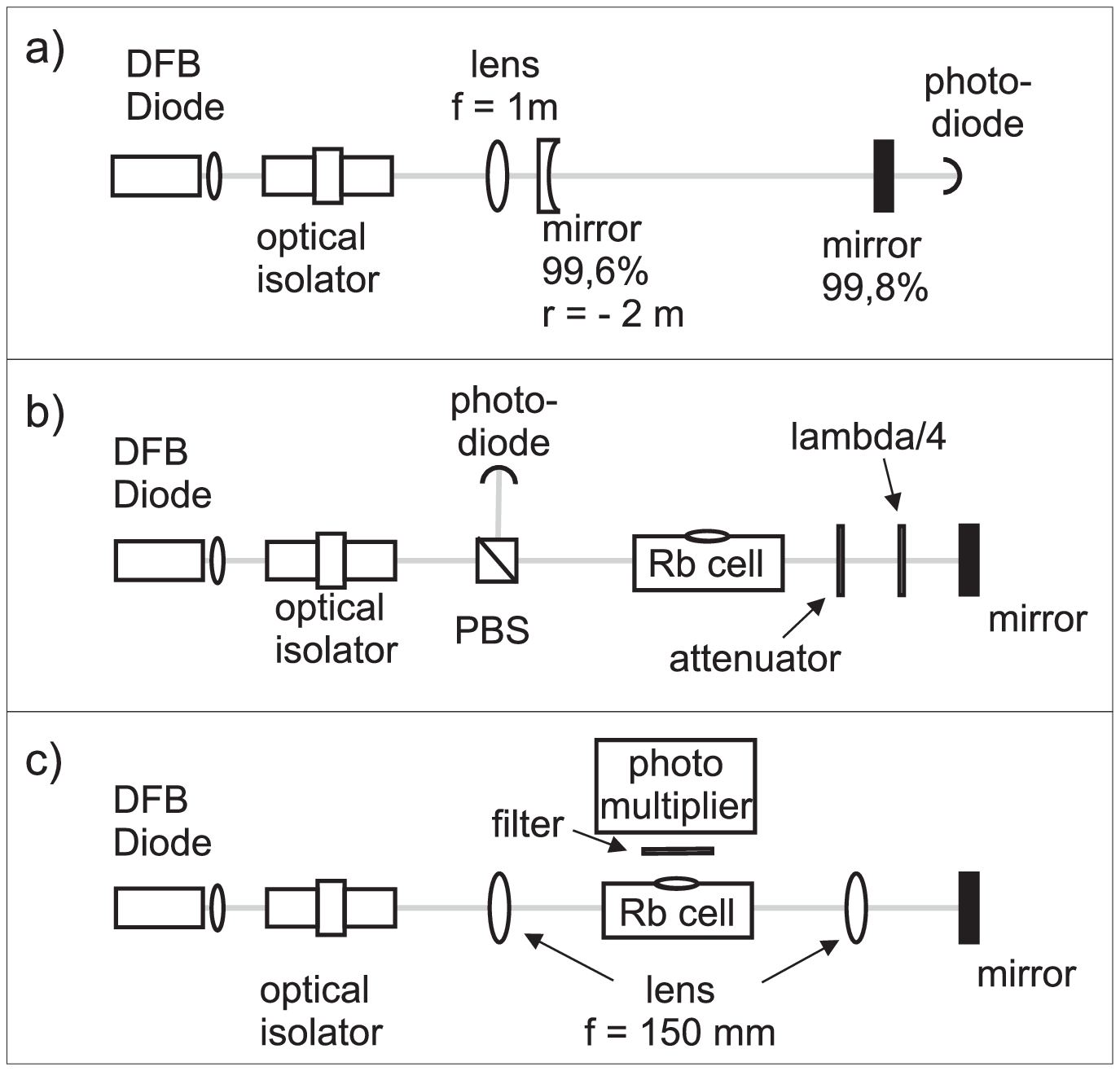}
\caption{Experimental setup of the experiment. The beam of the DFB
laser is collimated and passes a 60 dB optical isolator. (a) line
width measurements were done using a Fabry-Perot cavity. (b) Setup
used for rubidium saturation spectroscopy. For 2-photon
spectroscopy (c), the laser is focused into the cell and a photo
multiplier is added for detection of the blue fluorescence light.
}
\label{setup}
\end{figure}

For line width measurements, we used the setup sketched in Fig.
\ref{setup}a. The laser beam was coupled into a Fabry-Perot cavity
consisting of two mirrors spaced apart at a distance of 90 cm. The
input coupler (nominal reflectivity 99.6\%) had a radius of
curvature of -2000 mm while the second mirror (nominal
reflectivity 99.8\%) was plane. Mode matching was accomplished by
a lens (f = 1000 mm) placed 10 cm in front of the input coupler.
The transmitted light was detected by a fast photo diode with a
band width of 40 MHz. For saturation spectroscopy of the rubidium
D$_2$ line, the setup was modified as shown in Fig. \ref{setup}b.
The laser beam successively passed a polarizing beam splitter
(PBS), a glass cell filled with rubidium vapor, an attenuator
(attenuation factor = 6), and a quarter wave retardation plate
before being retro reflected. The probe beam, having passed the
retardation plate twice, was coupled out at the PBS and detected
by a fast photo diode.

For measurements of the 5S-5D two-photon transition (Fig.
\ref{setup}c), the attenuator and the retardation plate were
removed and the beam was focussed and re-collimated by two lenses
(f = 150 mm, distance 300 mm). The rubidium cell, placed at the
focus, was heated to $\sim100$ $^\circ $C in order to increase the
vapor pressure  and the 420 nm fluorescence light from the
radiative cascade 5D-6P-5S\cite{ryan1993,nez1993} was detected
with a photo multiplier.

\begin{figure}
\includegraphics[width=\columnwidth]{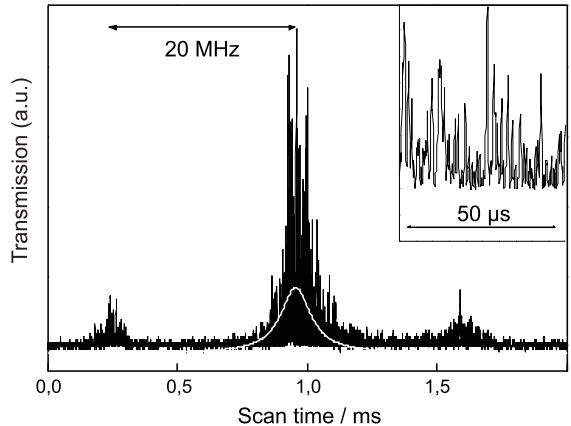}
\caption{Transmission spectrum of the Fabry-Perot cavity. The
laser wavelength was scanned at 30 MHz/ms. The spectrum consists
of the main resonance and two sidebands produced by 20 MHz
rf-modulation.  The inset shows an enlarged view of the main
resonance.}
\label{cavity}
\end{figure}

For the line width measurement of the DFB laser, the laser beam was
coupled into the aforementioned Fabry-Perot cavity (free spectral
range 166 MHz, estimated finesse $> 500$).  The transmitted light
was focussed onto the fast photo diode. In order to provide a
frequency reference, side bands were generated by applying a 20 MHz
sinusoidal modulation to the laser current by means of a Bias-Tee.
In addition the operating current was linearly modulated in order to
scan the laser across the resonance of the Fabry-Perot cavity. Fig
\ref{cavity} shows the transmitted signal
versus time. The velocity of the scan can be determined from the
position of the sidebands. In the given example it amounts to 30
MHz/ms. The 70 $\mu$s time interval for scanning across the
resonance corresponds to a laser line FWHM of 2 MHz. However, the
cavity transmission does not form a smooth resonance but rather
consists of several sharp needles with an individual width of
typically 0.5 $\mu$s each. These needles were also observed when the
laser frequency was scanned without additional rf modulation. We
have investigated the width of these needles for different scanning
rates between 30 MHz/ms and 2 GHz/ms. We observed that independent
of the scan speed, the width of the resonance envelope remained
constant in the frequency domain (2 MHz, line in fig. \ref{cavity})
while the width of the needles remained unchanged in the time domain
(0.5 $\mu$s). We then varied the laser current and found a strong
influence on the line width. Starting with a rather broad (8 MHz)
line at 45 mA (output power P = 10 mW) the line width decreased with
increasing current and reached a minimum for values between 60 and
80 mA (P = 35 mW). At higher currents the line width increased
again. We thus conclude that the observed line width of the envelope
is a technical line width, which results from frequency \lq
jumps\rq\ of a narrow laser line. The jumps may be due to spatial
hole burning, which leads to inhomogeneities in the charge carrier
density and consequently to mode fluctuations
\cite{pan1990,wenzel1991}. The observed constant needle width in the
time domain is consistent with a model for the laser spectrum that
consists of a very narrow frequency component which randomly sweeps
across the cavity resonance in a time interval shorter than the
cavity filling time. The observed width of 0.5 $\mu$s corresponds to
a cavity line width of 300 kHz which agrees reasonably well with the
expected line width of 200 kHz. This value also defines an upper
limit for the laser spectrum at time intervals shorter than 0.5
$\mu$s. For intervals of up to 5 $\mu$s the frequency excursion of
the jumping component remains within 2 MHz as can be seen from the
envelope. From spectroscopic data (see below) one can infer that the
excursions always stay within a range of 4 MHz.

\begin{figure}
 \includegraphics[width=\columnwidth]{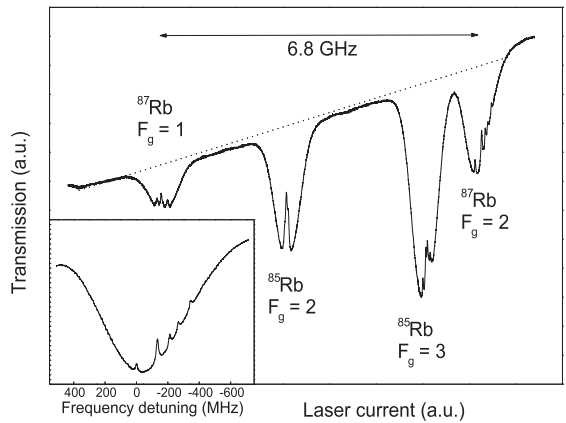}
\caption{Saturation spectrum of the D$_2$-line of $^{87}$Rb and
$^{85}$Rb. The figure shows the four Doppler broadened resonance
lines of the transitions  $^{87}$Rb $\mathrm{F_g} = 1$, $^{85}$Rb
$\mathrm{F_g} = 2$, $^{85}$Rb $\mathrm{F_g} = 3$, and $^{87}$Rb
$\mathrm{F_g} = 2$ (from left to right). The dotted line indicates
the intensity modulation resulting from a modulation of the diode's
driver current. The inset shows a close-up of  the $\mathrm{F_g}=2$
transition of $^{87}$Rb. All six Doppler-free peaks are resolved
(from left to right: $\mathrm{F_g} =2 \rightarrow \mathrm{F_e} = 3$,
cross-over $\mathrm{F_e}= 2\& 3$, cross-over $\mathrm{F_e} =3\& 1$,
$\mathrm{F_e}=2$, cross-over $\mathrm{F_e} =2\& 1$, and
$\mathrm{F_e}=1$).} \label{rb-spectra}
\end{figure}

In order to test the laser in a spectroscopic application, we
recorded the saturation spectra of the rubidium D$_2$ line (Fig
\ref{rb-spectra}). The intensities of pump and probe beam were 5
mW and 150 $\mu$W respectively, the beam diameters were
approximately 1mm. The spectra were acquired by scanning the
current of the diode. The current variation of 6 mA resulted in a
corresponding intensity change of 10\% within the scan range of
8.4 GHz. The laser power varied nearly proportional with the
current as shown in the figure. The four Doppler broadened lines
correspond to excitation of the two rubidium isotopes
$^{85}\mathrm{Rb}$ and $^{87}\mathrm{Rb}$ with two hypefine ground
states each: $^{87}\mathrm{Rb}$ ($\mathrm{F_g} = 1$),
$^{85}\mathrm{Rb}$ ($\mathrm{F_g} = 2$), $^{85}\mathrm{Rb}$
($\mathrm{F_g} = 3$), and $^{87}\mathrm{Rb}$ ($\mathrm{F_g} = 2$).
Their half widths of 700 MHz reflect the gas temperature of 300K.
Each resonance features several Doppler-free peaks, as shown in
the inset of Fig. \ref{rb-spectra} for the $^{87}\mathrm{Rb}$
$\mathrm{F_g} = 2$ transition. The measured line width of the
hyperfine transitions and crossover-resonances is 9 MHz, which is
close to the natural line width of 6 MHz.

A lock-in servo loop was used to stabilize the laser frequency to
Doppler-free D$_2$-line transitions from the $^{87}\mathrm{Rb}$
($\mathrm{F_g}=2$) ground state. The DFB laser was successively
locked to the transition $\mathrm{F_g} = 2 \rightarrow
\mathrm{F_e} = 3$ and the two cross-over resonances $\mathrm{F_g}
= 2 \rightarrow \mathrm{F_e} = 2\& 3$ and $\mathrm{F_g} = 2
\rightarrow \mathrm{F_e} = 1\& 3$ while the wavelength was
recorded with the wavelength meter. Table \ref{tab1} compares
\begin{table*}\small
 \begin{tabular}{|l|l|l|l|l|l|l|}
  \hline\noalign{\smallskip}
  Transition & Av. time & Literature value  & Measured mean  & Difference  & Freq STD & Relative   \\
    & [min:s] & [THz] & [THz] & [MHz] & [MHz] & stability  \\
  \noalign{\smallskip}\hline\noalign{\smallskip}
  $\mathrm F = 2 \rightarrow F' = 3$ & 4:38  & 384.228 115 2 & 384.228 127 8 & 12.6 & 1.3 & $3.5\cdot 10^{-9}$ \\
  $\mathrm F = 2 \rightarrow F' = 2 \& 3 $ & 30:27 & 384.227 981 9 & 384.227 992 9 & 11.0 & 2.4 & $6.4 \cdot10^{-9}$ \\
  $\mathrm F = 2 \rightarrow F' = 1 \& 3 $  & 5:21 & 384.227 903 4 & 384.227 912 4 & 8.0 & 1.6 & $4.3 \cdot10^{-9}$ \\
  \hline
\end{tabular}
\caption{Frequency-locking of a DFB laser to Doppler-free resonance
lines of $^{87}$Rb, transition 5S$_{1/2} \rightarrow$ 5P$_{3/2}$.
Literature values are taken from \cite{steck}. The time interval
over wich the laser frequency has been averaged is shown in the
column labeled \lq Av. time\rq .\lq Measured mean\rq\ denotes the
average value measured by the wavelength meter while the laser was
locked to the respective transition. The differences between the
literature reference and our measurement are shown in the fifth
column. \lq Freq STD\rq\ gives the measured standard deviation of
the laser frequency while the laser was in lock. The last column
states the relative frequency stability for the different lock
periods.} \label{tab1}
\end{table*}
precision data taken from literature \cite{steck} with our recorded
values. The deviation of the measured frequencies from the
literature values of approximately 10 MHz is well below the accuracy
of the wavelength meter which was calibrated with a stabilized
helium-neon laser at 632,99 nm. With the laser being locked to the
respective resonances, values between 1.4 and 2.5 MHz were measured
for the standard deviation of the frequency. This translates into a
relative frequency stability of 3-6 parts in $10^9$. However, as the
resolution of the utilized wavelength lies within the same range,
the actual frequency stability can be assumed to be even better.

\begin{figure}
\includegraphics[width=\columnwidth]{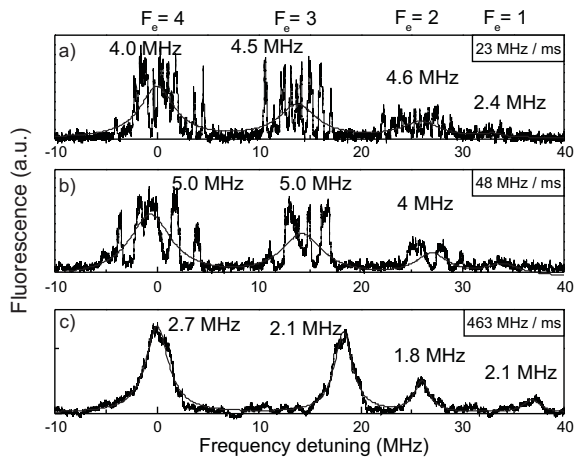}
\caption{Hyperfine spectrum of the two-photon transition
5S$_{1/2}$ ($\mathrm{F_g} = 2$) to 5D$_{5/2}$ ($\mathrm{F_e} = 4$
to 1) of $^{87}$Rb. The graph shows spectra acquired with
different scan speeds of 23 MHz/ms, 48 MHz/ms, and 463 MHz/ms,
respectively. At slow scan frequencies, the resonance lines
broaden and are successively decomposed into multiple peaks
resulting from high-frequent wavelength hops of the DFB diode.
Also shown are best-fit curves of the data. The line width
determined from a Lorentzian fit is given for each resonance. }
\label{twophoton}
\end{figure}

The 5S-5D two-photon transition of rubidium offers a narrow line
width ($\sim$500 kHz) \cite{nez1993} due to the long lifetime of
the 5D state. To probe this transition we focussed the laser beam
into the spectroscopy cell and modulated the wavelength by
scanning the operating current. Fig. \ref{twophoton} shows the
fluorescence signal, measured
with a photo multiplier, after excitation of the $\mathrm{5S_{1/2}},
\mathrm{F_g} = 2 \rightarrow \mathrm{5D_{5/2}}$ transition of
$\mathrm{^{87}Rb}$. Spectra a) - c) were recorded with scan speeds
of 23 MHz/ms, 48 MHz/ms and 463 MHz/ms, respectively. Also shown is
a fit curve of the four Lorentzian line profiles. The x-axis of the
plot is scaled to the frequency detuning using the frequency
separation between the $\mathrm{F_e = 4}$ and $\mathrm{F_e = 2}$
peaks. At slow scan speeds (a), the resonance lines are decomposed
into multiple peaks due to jitter of the laser frequency. As
observed in the Fabry-Perot spectrum (Fig. 5), the width of each
needle within the time domain does not depend on the scanning speed.
By scanning faster (b), the width of the line envelope remains
unchanged, while the number of needles per resonance decreases. At
even faster scan speeds (c), each resonance consists of only one
needle. The frequency hops now result in a displacement of the line
in frequency space. The fit curves of the peak structure in Fig.
\ref{twophoton} can be employed to investigate the technical line
width of the DFB diode as a function of scan velocity. The minimum
FWHM width of the measured lines is approximately 1.5-2 MHz. This
value is broader than the natural line width of 500 kHz, and agrees
with the previous measurements with the Fabry-Perot cavity (Fig.
\ref{cavity}). At even higher scan speeds, the width of the spectral
signatures appears to be Fourier limited, leading to line
re-broadening.

In summary, it can be stated that the type of laser described here
is very suitable for almost all standard high resolution
applications in alkali spectroscopy including laser cooling and
optical manipulation of ultra cold atoms. The high output power
allows for efficiently driving nonlinear processes and the spectral
width of the bare chip lies well within the natural line width of
the D$_1$ and D$_2$-transitions. The relatively low frequency noise
of the laser spectrum probably allows for further reducing the
spectral width to well below 100 kHz with standard servo loops that
link the injection current to an error signal derived from an atomic
spectrum or a stable resonator. The absence of any critical
mechanical or optical components in the laser's control system
permits a very reliable and stable operation. Together with its
extremely large mode-hop free tuning range these properties will
probably allow for the construction of compact and sophisticated
optical radiation sources that transfer the reliability and
performance known from radiofrequency technology into the optical
domain.


\begin{thebibliography}{}
\bibitem{wieman1991}
C.E. Wieman, L. Hollberg: Rev. Sci. Instrum. \textbf{62,}  1
(1991)
\bibitem{macadam1992}
K.B. MacAdam, A. Steinbach, C. Wieman: Am. J. Phys. \textbf{60,}
 1098 (1992)
\bibitem{ricci1995}
L. Ricci, M. Weidem\"uller, T. Esslinger, A. Hemmerich, C.
Zimmermann, V. Vuletic, W. K\"onig, T. W. H\"ansch: Opt. Comm.
\textbf{117,} 541 (1995)
\bibitem{tohmori1993}
Y. Tohmori, F. Kano, H. Ishii, Y. Yoshikuni, Y. Kondo: Elect.
Lett. \textbf{29,} 1350 (1993)
\bibitem{labachelerie1994}
M. de Labachelerie, K. Nakagawa, M. Ohtsu: Opt. Lett. \textbf{19,}
 840 (1994)
\bibitem{poulin1994}
M. Poulin, C. Latrasse, M. T\^{e}tu, M. Breton: Opt. Lett.
\textbf{19,} 1183 (1994)
\bibitem{bickleder}
G. Bickleder, A. Zach, W. Kaenders, Patent DE 199 26 801 (2000)
\bibitem{ryan1993}
R.E. Ryan, L.A. Westling, H.J. Metcalf: J. Opt. Soc. Am. \textbf{B
10,} 1643 (1993)
\bibitem{nez1993}
F. Nez, F. Biraben, R. Felder, Y. Millerioux: Opt. Com.
\textbf{102,} 432 (1993)
\bibitem{pan1990}
X. Pan, H. Olesen, B. Tromborg: IEEE Photon. Tech. Lett. \textbf{2,}
312 (1990)
\bibitem{wenzel1991}
H. Wenzel,  H.J. W\"unsche, U. Bandelow: Electron. Lett.
\textbf{27,}  2301 (1991)
\bibitem{steck}
D.A. Steck http://steck.us/alkalidata, Revision 1.6 (2003)
\end{thebibliography}
\end{document}